# Phonon-limited electrical transport properties of intermetallic compound YbAl$_3$ from first-principles calculations

Jinghua Liang[1], Dengdong Fan[1], Peiheng Jiang[1], Huijun Liu[1,*], Wenyu Zhao[2]

[1]*Key Laboratory of Artificial Micro- and Nano-Structures of Ministry of Education and School of Physics and Technology, Wuhan University, Wuhan 430072, China*

[2]*State Key Laboratory of Advanced Technology for Materials Synthesis and Processing, Wuhan University of Technology, Wuhan 430070, China*

We combine first-principles calculations and Boltzmann transport theory to study the electrical transport properties of intermetallic compound YbAl$_3$. To accurately predict the electronic relaxation time, we use the density functional perturbation theory and Wannier interpolation techniques which can effectively treat the electron-phonon scattering. Our calculated transport coefficients of YbAl$_3$ are in reasonable agreement with the experimentally measured results. Strikingly, we discover that in evaluating the Seebeck coefficient of YbAl$_3$, the scattering term has a larger contribution than the band term and should be explicitly considered in the calculations, especially for the case with localized bands near the Fermi level. Moreover, we demonstrate that by reducing the sample size to less than ~30 nm, the electronic thermal conductivity of YbAl$_3$ can be sufficiently suppressed so that the thermoelectric figure of merit can be further enhanced.

Thermoelectric (TE) materials can directly convert heat into electricity and vice versa by utilization of the Seebeck effect for power generation and the Peltier effect for cooling. Due to their advantages of small size, good stability, no noise and no pollution, TE devices have attracted much attention in recent years. The efficiency of TE materials is usually described by the dimensionless figure of merit $ZT = S^2 \sigma T / \kappa$, where $S$, $\sigma$, $\kappa$, $T$ are the Seebeck coefficient, the electrical conductivity, the thermal conductivity, and the absolute temperature, respectively. Good TE materials have larger $ZT$ values which requires higher power factor ($S^2\sigma$) and lower thermal conductivity $\kappa$ (including the electronic part $\kappa_e$ and the lattice part $\kappa_L$). Most TE materials are semiconductors, since they have moderate power factor and relatively lower thermal conductivity simultaneously. However, it is found that some intermetallic compounds, such as YbAl$_3$, can also exhibit good thermoelectric performance. For example, the measured Seebeck coefficient of YbAl$_3$ can be as high as about −80 μVK$^{-1}$ at room temperature [1, 2], which is unusually large for metallic systems. Such large value of Seebeck coefficient is thought to be originated from the *f*-electron of Yb which leads to a sharp density of states (DOS) around the Fermi level. As a consequence, YbAl$_3$ could exhibit much higher power factor which outperformances those of the state-of-the-art Bi$_2$Te$_3$-based TE materials [3, 4].

---

* Author to whom correspondence should be addressed. Electronic mail: phlhj@whu.edu.cn

However, the $ZT$ value of YbAl$_3$ is still limited by its relatively large thermal conductivity, which could be reduced by doping B, C [5] or Mn [6] into the YbAl$_3$ lattice as interstitial atom, by substituting Yb with Sc [7], Er and Lu [8], or Al with Sn [9]. Introducing second phase [10] or excessive Al [11] in the preparation of YbAl$_3$ has been also suggested as possible strategy. However, in the experiments, it is difficult to fabricate YbAl$_3$-based solid solution due to the peritectic nature of Yb-Al phase diagram and no phase width of YbAl$_3$ [12, 13]. It is well known that nanostructuring could significantly enhance the thermoelectric performance [14], which is however not considered so far for the intermetallic compound YbAl$_3$. In this work, first-principles calculations and Boltzmann transport theory are combined to study the electrical transport properties of YbAl$_3$, where the electron relaxation time is accurately predicted by using the electron-phonon Wannier interpolation techniques. It is surprising to find that the scattering term of the Seebeck coefficient has a larger contribution than the band term, which is quite different from that predicated by using constant relaxation time approximation (CRTA). We also demonstrate how the electronic thermal conductivity of YbAl$_3$ can be effectively reduced by nanostructuring so that enhanced $ZT$ could be expected.

Our first-principles calculations are performed within the framework of density functional theory (DFT), as implemented in the so-called QUANTUM ESPRESSO package [15]. We use the norm-conserving pseudopotential and the exchange-correlation functional is in the form of Perdew-Burke-Ernzerhof (PBE) [16]. The energy cutoff is set as 60 Ry and the atomic positions are fully relaxed. Based on the calculated band structure, the electronic transport coefficients of YbAl$_3$, e.g., the electrical conductivity $\sigma$, the Seebeck coefficient $S$, and the electronic thermal conductivity $\kappa_e$, can be obtained from the Boltzmann transport theory [17]. Here, the electron-phonon relaxation time $\tau_{ep}$ of YbAl$_3$ is accurately predicted by using the density functional perturbation theory (DFPT) and the Wannier interpolation techniques [18]. The electron-phonon coupling matrix element is first calculated on a coarse 8 × 8 × 8 **k**-mesh and a 4 × 4 × 4 **q**-mesh, and then interpolated to an ultra-dense 50 × 50 × 50 **k**-mesh and a 50 × 50 × 50 **q**-mesh via the maximally localized Wannier functions, as implemented in the electron-phonon Wannier (EPW) package [19, 20].

The YbAl$_3$ compound crystallizes in the AuCu$_3$ structure with the space group of Pm$\overline{3}$m (No. 221), where Yb atoms occupy the 1$a$ (0, 0, 0) site and Al atoms occupy the 3$c$ (0, 0.5, 0.5) site. The optimized lattice constant is 4.20 Å, which agrees well with previous theoretical and experimental results [21, 22]. Figure 1 plots the band structure and DOS of YbAl$_3$. It is clear to find that the 4$f$ states of Yb are well localized at about 0.3 eV below the Fermi level, which results in very sharp DOS and thus large Seebeck coefficient. According to the Mott expression, the Seebeck coefficient $S$ is related to the logarithmic energy derivatives of DOS $N(\varepsilon)$ and the electronic relaxation time $\tau(\varepsilon)$ at the Fermi level $\varepsilon_F$ [23, 24]:

$$S = -\frac{\pi^2}{3}\frac{k_B^2 T}{e}\left[\frac{\partial \ln N(\varepsilon)}{\partial \varepsilon} + \frac{\partial \ln \tau(\varepsilon)}{\partial \varepsilon}\right]_{\varepsilon_F} = S_N + S_\tau \qquad (1)$$

where $k_B$ is the Boltzmann constant and $e$ is the electron charge. Note we have separated the Seebeck coefficient into the so-called band term $S_N$:

$$S_{\mathrm{N}} = -\frac{\pi^2}{3}\frac{k_B^2 T}{e}\frac{\partial \ln N(\varepsilon)}{\partial \varepsilon}\bigg|_{\varepsilon_F} \qquad (2)$$

and the scattering term $S_\tau$:

$$S_\tau = -\frac{\pi^2}{3}\frac{k_B^2 T}{e}\frac{\partial \ln \tau(\varepsilon)}{\partial \varepsilon}\bigg|_{\varepsilon_F} \qquad (3)$$

Indeed, we find from Equation (2) that a sharp DOS around the Fermi level can lead to the large band term of the Seebeck coefficient, which was confirmed by experimental measurements and further support the idea that good thermoelectric can be found in materials with a sharp singularity in the DOS close to the Fermi level [25].

When dealing with the electrical transport properties, the relaxation time should be carefully treated. In contrast to the generally used CRTA, here we consider the energy dependence of the relaxation time by calculating the electron-phonon coupling matrix elements using the DFPT and Wannier interpolation techniques [18, 19, 20]. In Figure 2, we show the room temperature relaxation time $\tau_{ep}$ of YbAl$_3$ plotted as a function of energy. We see that the overall value of $\tau_{ep}$ is inversely proportional to the DOS shown in Fig. 1(b). This is reasonable since the electron-phonon scattering rate ($1/\tau_{ep}$) is proportional to the phase space available (i.e. the DOS) for carrier scattering [18, 26]. Moreover, we find that $\tau_{ep}$ changes rapidly around the Fermi level (indicated by red arrow in Fig. 2). According to Equation (3), this suggests that the scattering term $S_\tau$ can also contribute significantly to the Seebeck coefficient as compared with the band term $S_N$. It should be mentioned that the scattering term $S_\tau$ is usually neglected in the evaluation of Seebeck coefficient, especially when the CRTA is aspoted. We will come back to this point later.

By inserting the calculated relaxation time into the Boltzmann equations, we can evaluate the electrical conductivity $\sigma$, the Seebeck coefficient $S$, the electronic thermal conductivity $\kappa_e$ and thus the $ZT$ value of YbAl$_3$, which are all shown in Figure 3 as a function of temperature. We see that the electrical conductivity $\sigma$ shows typical metallic behavior, that is, it decreases with increasing temperature. At room temperature, the calculated electrical conductivity is $3.0\times10^6$ Sm$^{-1}$, and the Seebeck coefficient is $-71.2$ μVK$^{-1}$, both of which are in reasonable agreement with experimental results [5, 6, 7, 8, 9, 10, 11]. As a result, the room temperature power factor of YbAl$_3$ reaches $1.5\times10^{-2}$ Wm$^{-1}$K$^{-2}$, which is much higher than that of Bi$_2$Te$_3$-based thermoelectric materials [3, 4]. However, since the calculated electronic thermal conductivity $\kappa_e$ is as high as 24.4 Wm$^{-1}$K$^{-1}$ at room temperature, the corresponding $ZT$ value of YbAl$_3$ is only 0.15. Note here an average experimental value of lattice thermal conductivity of 6.0 Wm$^{-1}$K$^{-1}$ [9, 10, 11] is used to calculate the $ZT$ value. To further enhance the thermoelectric performance of YbAl$_3$, major efforts should be devoted to reduce its relative large electronic thermal conductivity.

The enhanced Seebeck coefficient of YbAl$_3$ is often accounted by a model [24] with renormalized narrow $f$-electron band of the Lorenzian shape, which only considers the band term $S_N$ and ignores the scattering term $S_\tau$. It should be also noted the CRTA is generally adopted [27, 28, 29] in calculating the Seebeck coefficient, which implies that only the band term $S_N$ is considered. To illustrate the vital importance of the scattering term $S_\tau$, we show in Figure 4 the room temperature Seebeck coefficients of YbAl$_3$ as a function of the chemical potential $\mu$, where the results calculated with the energy dependent relaxation time $S(\tau_{ep})$ and that under CRTA $S(\text{CRTA})$ are both shown. According to Eq. (1), $S(\text{CRTA})$ is just the band term $S_N$, and the scattering term can be directly obtained by $S_\tau = S(\tau_{ep}) - S_N$. At the Fermi level ($\mu = 0$ eV), the calculated $S(\tau_{ep}) = -71.2$ μVK$^{-1}$, $S_N = -27.6$ μVK$^{-1}$, and $S_\tau = -43.6$ μVK$^{-1}$. It is striking to see that the absolute value of scattering term $S_\tau$ is even larger than that of the band term $S_N$. In fact, we see from Fig. 4 that when the chemical potential is close to the Yb-$f$ states, the Seebeck coefficient calculated with energy dependent relaxation time is always much larger than that under the CRTA. Our theoretical results thus demonstrate that the scattering term cannot be neglected in the calculation of the Seebeck coefficient, especially when there are localized bands around the Fermi level.

Compared with conventional semiconducting thermoelectric materials, the metallic $YbAl_3$ exhibit a relatively larger electronic thermal conductivity, which could be reduced by nanostructuring [14, 30]. In principle, the electrons can be efficiently blocked when their mean free path is longer than the characteristic length of scattering. To estimate the effectiveness of nanostructures, we show in Figure 5 the normalized accumulated contribution to the electronic thermal conductivity $\kappa_e$ from individual electrons with respect to their mean free path $\lambda$ at room temperature. In our calculations, we assume that electrons with mean free path longer than a given value are totally blocked so that they do not participate in the transport process. It can be seen from Fig. 5 that when the characteristic length scale of nanostructure is less than ~30 nm, the $\kappa_e$ of $YbAl_3$ can be effectively decreased. For example, if the sample size can be made as small as ~9 nm, the electronic thermal conductivity will be reduced by half compared with of its bulk value. Moreover, since the mean free path of phonons is usually larger than that of electrons, the lattice thermal conductivity of $YbAl_3$ can be also reduced. It is thus reasonable to expect that the $ZT$ value of $YbAl_3$ could be significantly enhanced by such kind of nanostructuring.

In summary, we have investigated the electrical transport properties of intermetallic compound $YbAl_3$ using first-principles calculations and Boltzmann transport theory. Due to the marked energy dependence of the relaxation time, we find that the system exhibit unusually large Seebeck coefficient with comparable band and scattering terms. Although the $ZT$ value of $YbAl_3$ is obviously lower than those of many good thermoelectric materials, the much higher power factor suggests its promising thermoelectric application provided that the electronic thermal conductivity can be effectively reduced by means such as nanostructuring.

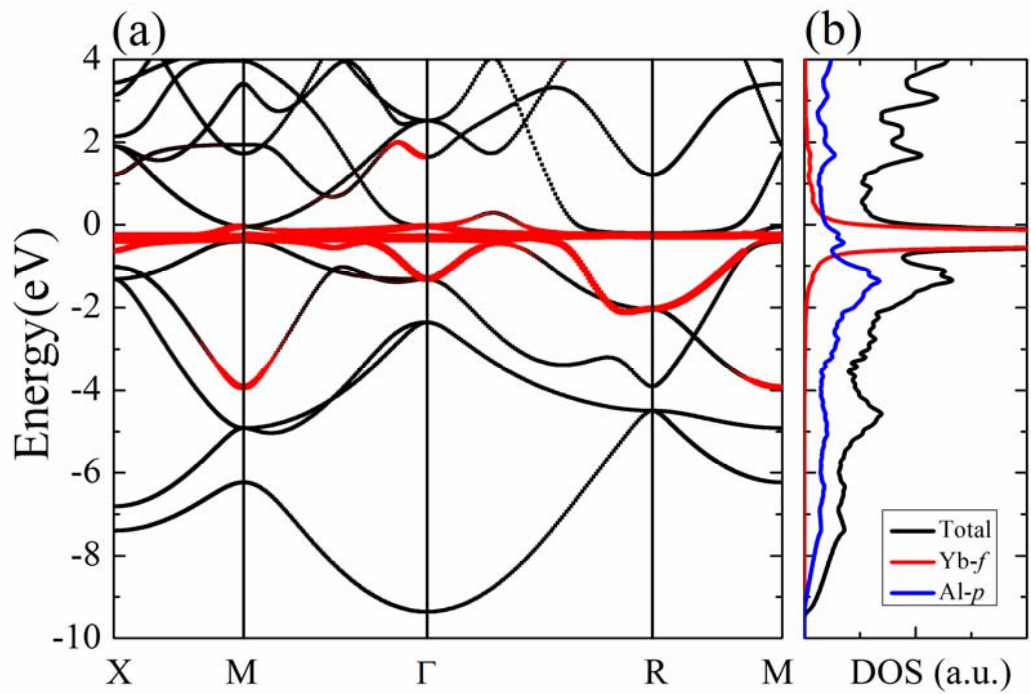

**Figure 1** The band structure (a) and density of states (b) of $YbAl_3$. The size of red dots in the energy bands is proportional to the contribution from *f* electrons of Yb atom.

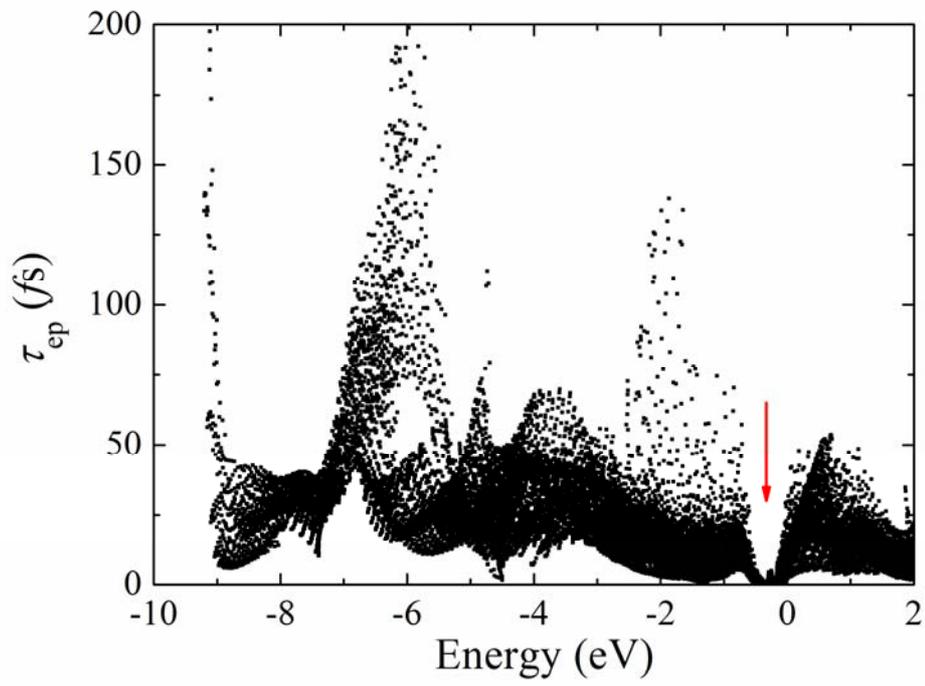

**Figure 2** The energy dependence of room temperature relaxation time of YbAl$_3$. The red arrow indicates a deep valley around the Fermi level.

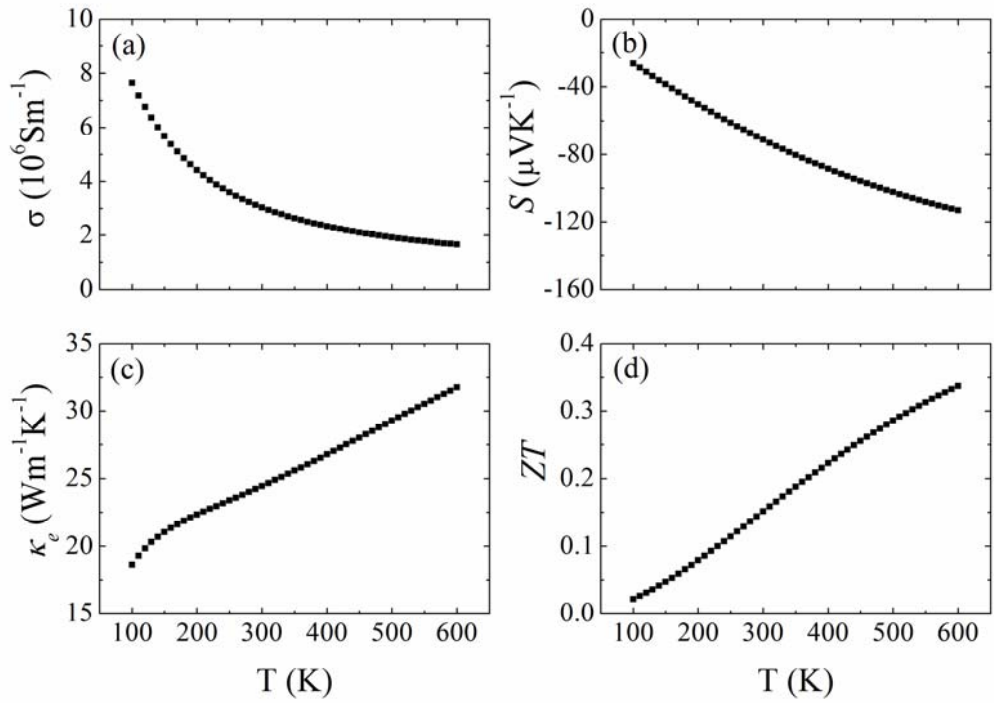

**Figure 3** The calculated temperature dependence of the electrical conductivity $\sigma$ (a), the Seebeck coefficient $S$ (b), the electronic thermal conductivity $\kappa_e$ (c), and the $ZT$ value (d) of YbAl$_3$.

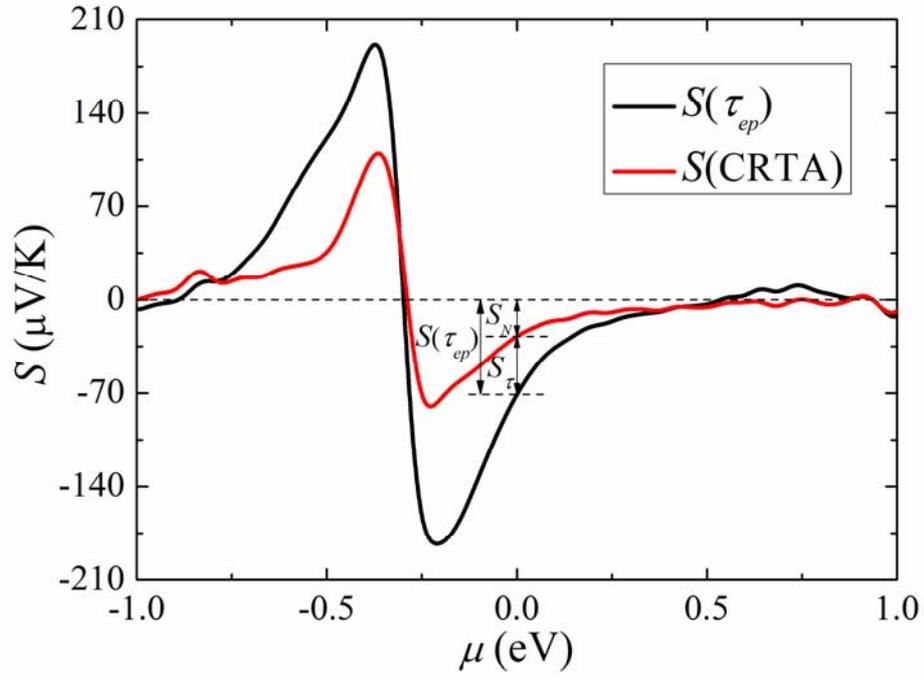

**Figure 4** The Seebeck coefficient of YbAl$_3$ calculated with energy dependent relaxation time $S(\tau_{ep})$ (black line) and that under the constant relaxation time approximation $S(\text{CRTA})$ (red line), plotted as a function of the chemical potential $\mu$ at room temperature.

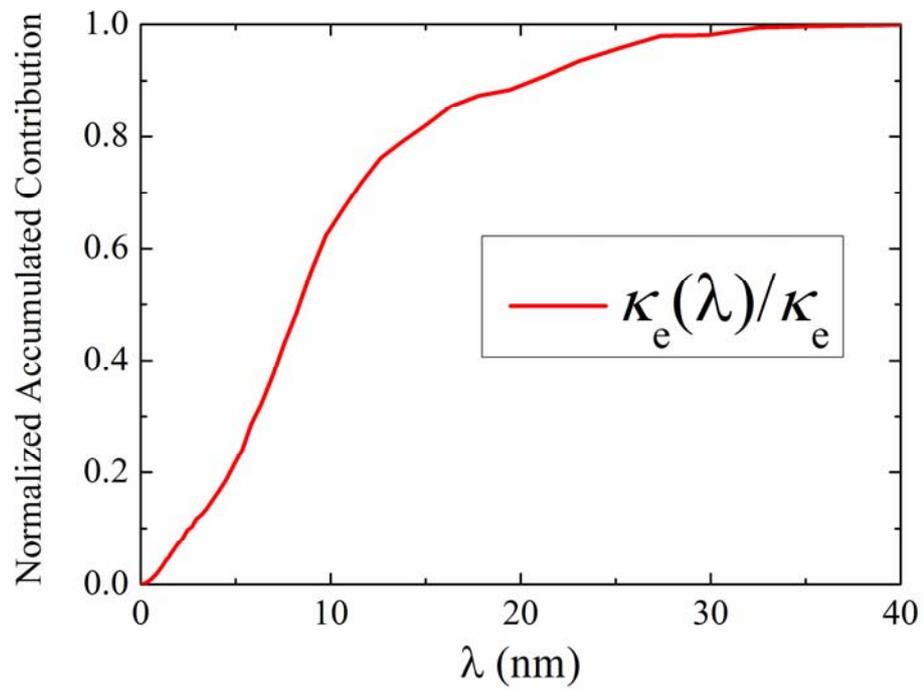

**Figure 5** Normalized accumulated contribution to the electronic thermal conductivity $\kappa_e(\lambda)/\kappa_e$ from individual electrons with respect to their mean free path $\lambda$.